\documentclass{aa}
\usepackage{graphicx}
\begin{document}
\title{Reddening map of the Large Magellanic Cloud bar region
}
\author{Annapurni Subramaniam}
\institute{Indian Institute of Astrophysics, II Block, Koramangala, Bangalore
560034, India}
\offprints{Annapurni Subramaniam, e-mail: purni@iiap.ernet.in}

\date{Received / accepted}

\authorrunning{Subramaniam}
\titlerunning{Reddening map of the LMC bar region}

\abstract{E(V$-$I) reddening values for 1123 locations in the bar region of the Large
Magellanic Cloud are presented.
V, I photometry of red clump stars identified in the Optical Gravitational 
Lensing Experiment II catalogue of LMC were used to estimate reddening.
E(V$-$I) values were estimated as the difference between observed and the  
characteristic values of (V$-$I) for the red clump population in a given region. It is found that most
of the regions in the bar
have reddening values less than 0.1 mag with only a few locations having values more than 0.2 mag.
Eastern side of the bar is found to be more reddened  when compared to the western side,
with similar and relatively small values for differential reddening
as in most parts of the bar. Increased reddening in the eastern end of the bar
could be caused by a small fraction of the H I clouds in the line of sight. 
A high density of HI clouds located in the eastern end of the bar should
have caused very high reddening in these regions, whereas only a relatively
small increase in the reddening is estimated.
This indicates that most of the H I clouds in this direction are likely to be 
located behind the bar.
}

\maketitle

\section{Introduction}
The Large Magellanic Cloud (LMC) is the nearest galaxy where one can resolve individual stars
and study them. Since metallicity of the LMC is about one third of our Galaxy,
dependence of various processes on metallicity can be understood by studying
stars in various evolutionary phases in the LMC. 
One of the main hurdles in deriving the observed characteristics is in estimating
reddening in the line of sight towards them. Properties
of similar stars located in different parts of the LMC can be compared only if 
reliable estimation of reddening is available for various locations. 
Thus a map of reddening with sufficient spatial resolution is necessary.
In recent years, the LMC has been very thoroughly studied using various surveys,
for example, Optical Gravitational Lensing Experiment II (OGLE II Udalski et al. \cite{u2000}), 
Magellanic Clouds Photometric Survey 
(MCPS, Zaritsky et al. \cite{z97}), Massive Compact Halo Objects (MACHO) survey 
(Alcock et al. \cite{a00}).
These surveys can be used to map reddening in the LMC. 
Since data from a survey catalogue are homogeneous, systematic error in the reddening 
estimation from such a catalogue will be minimum. 

Some of the previous efforts to estimate reddening using survey catalogues are listed
below. Sumi (\cite{su04}) estimated E(V$-$I) reddening and extinction map towards
the Galactic bulge using the OGLE II data towards the bulge. Popowski et al. (\cite{pop03})
presented a reddening map of the Galactic bulge/bar based on (V$-$R) colours from the 
photometry of the MACHO microlensing survey. Wozniak \& Stanek (\cite{ws96}) used 
red clump stars in the Galactic bulge from the OGLE II survey to estimate reddening curve
and Stanek (\cite{s96}) used their method to obtain a high resolution reddening map of the
Baade's window.

A high resolution reddening map of the LMC bar region estimated using the OGLE II catalogue
is presented here.
In this study, we used red clump stars as a tracer to estimate reddening.
Red clump stars in the bar region of the LMC can be assumed to belong to a small
range of age and metallicity. Studies by 
Subramaniam \& Anupama (\cite{sa02}), 
Olsen \& Salyk (\cite{os02}) and
van der Marel \& Cioni (\cite{vc01})
found no noticeable change in the age or metallicity of red clump
population in the LMC. Hence red clump stars can be considered to be a homogeneous
population. This assumption is used to choose a characteristic
luminosity and colour for the population. 
The red clump stars are considered as standard candles and have been
widely used for estimation of distance to the LMC ( Stanek et al. \cite{szh98}, Udalski et al. 
\cite{u98}, Cole \cite{c98}). In these studies, 
characteristic luminosity of the population is used to estimate distance. 
In the present study, their characteristic colour is used to estimate reddening.
The estimates of reddening obtained here are average values of red clump stars
located in a given region. The relative change of this average value across the bar of the
LMC is a good indicator of the variation in reddening across the bar region. There may also
be differential reddening within LMC, for stars located in the same region. The error in
the estimation of reddening
is a proxy for the amount of differential reddening in that direction.
Reddening results of this analysis have been used by Subramaniam (\cite{as03}) to understand the
structure of the bar. Detailed reddening results 
are presented here so that they could be used for other studies.

\section{Data}
\begin{figure}
\resizebox{\hsize}{!}{\includegraphics{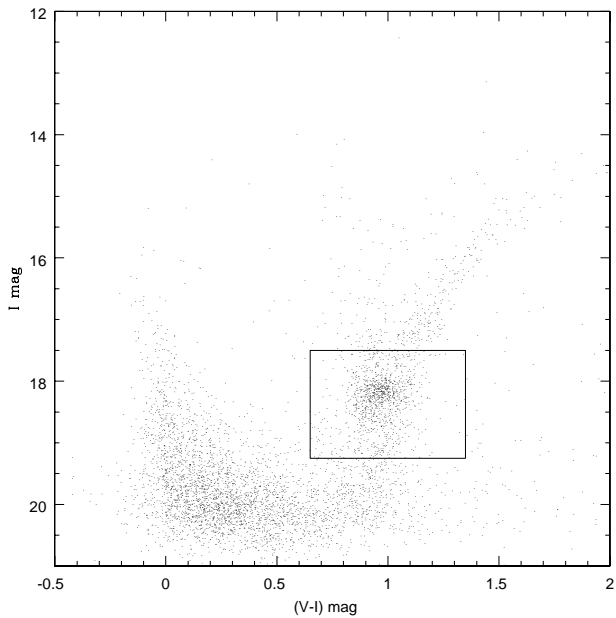}}
\caption{A typical I (mag) versus (V$-$I) (mag) CMD is shown. The red clump stars are identified
as stars within the box shown.}
\label{figure1}
\end{figure}
OGLE II survey (Udalski et al. \cite{u2000}) consists of photometric data of 7 million stars 
in B, V and I 
pass bands in the central 5.7 square degree of the LMC. The above catalogue presented 
data for 26 regions, of which 21 regions were studied here,
which are located within 2.5 degree from the optical center of the LMC 
(RA=$05^h19^m38.^s0$, Dec=$-69^o27'5."2$, de Vaucouleurs \& Freeman \cite{df73}).
In this study,
the total observed region was divided into 1344 sections. 
Each strip was divided into 4$\times$16 square sections, each having an 
area of 3.56$\times$3.56 arcmin$^2$. Thus each of the
21 strips was divided into 64 square sections, giving rise to 1344 sections.
For each section, an I versus (V$-$I) colour magnitude diagram (CMD) was constructed.
Red clump stars were identified using I vs (V$-$I) 
CMD, as stars within the box, as shown in figure 1. Blue and red boundaries
of the box in (V$-$I) are 0.65 mag and 1.35 mag respectively, whereas the bright
and faint boundaries in I mag are 17.5 mag and 19.25 mag respectively.
The position of red clump population
was found to be located within this box in almost all the locations and hence the location 
of this box was kept fixed in the CMD.
As the density of stars is not uniform in the bar, the number
of red clump stars also vary accordingly. The number of red clump stars identified was
found to vary between 800 -- 3000 with an average of 1500 red clump stars
per region.
The photometric data 
suffer from incompleteness in the data due to crowding effects, and 
incompleteness in the data in I and V pass bands are tabulated in Udalski et al. (\cite{u2000}).
Within the box of red clump stars, a two-dimensional grid was formed using a 
bin size of 0.010 mag in (V$-$I) and 0.025 mag in I magnitude. For each location in the grid,
there would be two values of incompleteness, one from V frame and the other from I frame.
The lower of the above two values was considered as the true incompleteness and applied
to obtain the actual number of stars. The selected value of incompleteness varies
within the box as well as with the location of the region in the LMC bar.
In crowded regions, minimum value of completeness was found to be 81\%, whereas
for a least crowded region, minimum value was 94\%. Above values refer to the
bottom-right corner of the box and also indicate the scaling adopted to
obtain the actual value of red clump stars from the observed number in the CMD.
Thus the estimated incompleteness factor for each region was applied and the total 
number of red clump stars were estimated for all the 1344 sections. 

\section{Estimation of Reddening map}
It is evident
from the CMD that red clump stars have a spread in luminosity and colour, within
the box in the CMD.
Number distribution of stars with respect to (V$-$I) colour shows a Gaussian like
distribution, with a peak and a width. 
The peak of the profile can be considered to be
due to major fraction of the red clump stars in the region considered.
This peak of the distribution would get shifted
towards redder colour due to reddening, with respect to its characteristic colour. 
Thus by estimating the shift in the peak of the distribution,
average reddening towards the location can be estimated. 
The spread in profile could be due to factors like,
differential reddening, presence of binaries, (real as well as line of sight) and 
difference in metallicity. 
The differential reddening is due
to difference in reddening towards stars located in a region in the same line of sight, because of
gas/dust located between the closest and the farthest red clump star in the line of sight.
If two stars are located in the same line of sight,
which cannot be resolved, then the estimated magnitude and colour are the integrated
values of the two stars. If we consider a star in the red clump and another star not in the
red clump, then these two stars have different intrinsic (V$-$I) and I magnitudes.
If these two stars happen to be in the same line of sight, then, the integrated values of the
I magnitude and (V$-$I) colour will not be the same as that of the red clump star. The resulting
change will depend on the location of the second star in the CMD. This is true for real binaries
also. The net effect is to produce a scatter around the oberved clump in the CMD, increasing the
width of the profile.

\begin{figure}
\resizebox{\hsize}{!}{\includegraphics{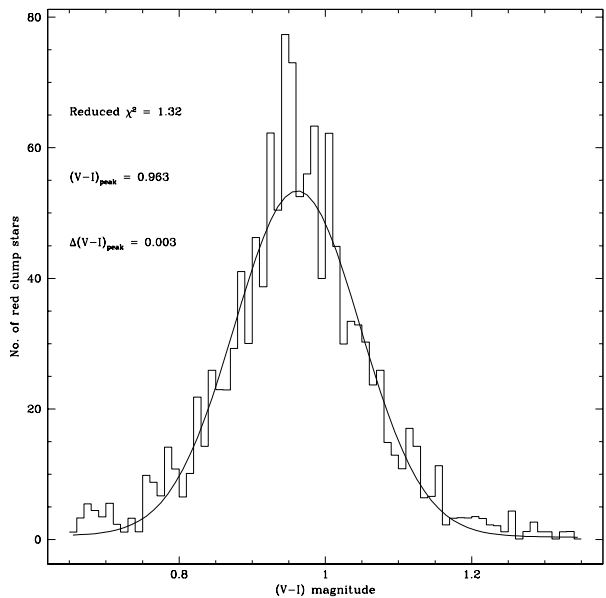}}
\caption{Number distribution of red clump stars as a function of (V$-$I) mag is shown
as a histogram. The fitted curve is also shown. The estimated values of peak of the function,
error in the estimation of the peak and reduced $\chi^2$ value of the fit are also shown
in the figure. This data is
obtained within a area of 3.56$\times$3.56 arcmin$^2$ centered at  
the location RA = 5$^h$ 32$^m$ 45$^s$ and Dec = $-$70$^o$ 25' 37''. }
\label{figure2}
\end{figure}
In order to estimate the number
distribution of red clump stars, stars were binned using a bin size of 0.010 mag.
The distribution was then fitted with a Gaussian + quadratic function. Similar
function was used by Udalski et al. (\cite{u99}) as well as Olsen \& Salyk (2002) to fit
red clump distribution. One such distribution for a location is shown in figure 2.
The histogram represents the observed number distribution and the curve represents the
best fit function. Estimated values of the peak, its error and goodness of fit
are also indicated in the figure. This procedure was repeated for all the 1344 sections.
Goodness of fit was estimated using reduced $\chi^2$ values and these values were found
to be similar for more than 90\% of the regions. Some regions showed poor fit along with
relatively higher values for width of the profile. These regions were not included
for reddening estimation. Average value of $\chi^2$ for the fit was found to be 1.39.
Regions with fit of the distribution worse than 2.5 were rejected and the number
of regions for reddening estimation was reduced to 1123. 
\begin{figure}
\resizebox{\hsize}{!}{\includegraphics{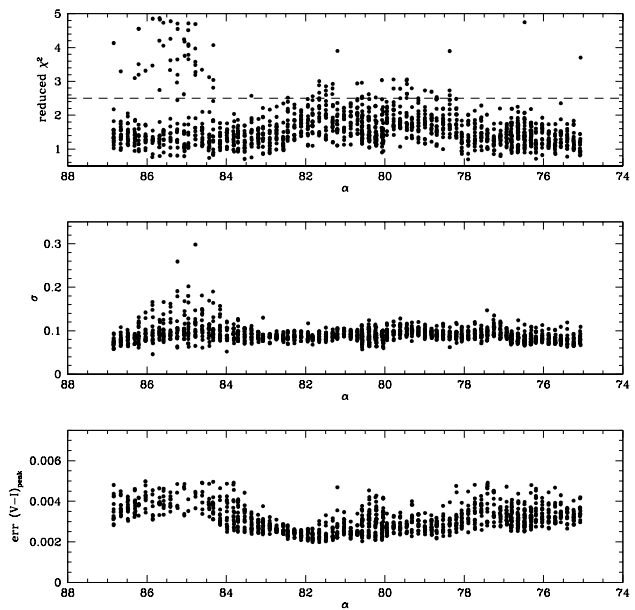}}
\caption{ The estimates of reduced $\chi^2$ value, width of the profile ($\sigma$) and
error in the peak value of (V$-$I) are shown as a function of $\alpha$.
In the top panel, the dashed line indicates the cut-off value of reduced $\chi^2$ value. 
}
\label{figure3}
\end{figure}

The distribution of reduced $\chi^2$, $\sigma$ and
error in the estimation of (V$-$I)$_{peak}$  are shown as a function of $\alpha$ in figure 3. 
Since the data spans about 12 degree in right ascension, $\alpha$,
and the span in declination, $\delta$ is only about 2 degrees,
the errors are shown as a function of $\alpha$. 
Reduced $\chi^2$ values were found to be mostly less than 2.0. Slightly higher values were found
between 78$^o$ and 82$^o$ and very high values were found eastward of 84$^o$. The values of $\sigma$
as shown in the middle panel indicates that higher values are seen only
in the eastern side. The error in (V$-$I)$_{peak}$ as shown in the lower panel, does not show
any trend between $\alpha$ = 75$^o$ -- 84$^o$. Locations eastward of 84$^o$ are found to show 
higher reddening and the maximum value of error was found to be 0.005 mag.
Significant number of locations between $\alpha$ = 84$^o$ -- 87$^o$ are found to show
poor fits. One of the reasons for the poor fit could be that there is considerable amount
of internal reddening within the LMC in these regions. This reason is supported by the fact 
that a large amount
of H I gas is found in the radio obervations in the eastern end of the bar. 
Subramaniam (\cite{as04}) found that between $\alpha$ = 84$^o$ -- 85$^o$, the red clump
stars have relatively large depth in the line of sight. The above two effects could
result in broadening of the peak and an increase in the value of $\sigma$. In general,
oberved profile could shift considerably from the fitting function, resulting in poor fits.

Peak values of (V$-$I) colour at each location was used to estimate
reddening. Reddening was calculated using the relation
 E(V$-$I) = (V$-$I)$_(obs)$ -- 0.92 mag.
The intrinsic colour of red clump stars was assumed to be 0.92 mag 
(Olsen \& Salyk \cite{os02}). Thus E(V$-$I) values were estimated for 1123 locations in the
LMC and are tabulated in Table 1. The complete table is available only electronically at the CDS
via anonymous ftp to cdsarc.u-strasbg.fr (130.79.128.5) or 
via http://cdsweb.u-strasbg.fr/cgi-bin/qcat?J/A+A/
and only a sample of Table 1 is presented here.
\begin{table}
\caption{A subsample of E(V$-$I) reddening values. The complete table is available through CDS.}
\begin{tabular}{llll}
\hline
$\alpha$ (J2000) & $\delta$ (J2000) & E(V$-$I) & error \\
(degrees) & (degrees) & (mag) & (mag) \\
\hline
  5  0 17&$-$69 20 34&   0.084&   0.004  \\
  5  0 17&$-$69 16 58&   0.085&   0.003  \\
  5  0 17&$-$69 24  7&   0.111&   0.004  \\
  5  0 17&$-$69 13 33&   0.058&   0.003  \\
  5  0 18&$-$69  6 28&   0.057&   0.003  \\
  5  0 18&$-$69 10  1&   0.069&   0.003  \\
  5  0 18&$-$69  2 52&   0.095&   0.004  \\
  5  0 19&$-$68 59 23&   0.088&   0.004  \\
  5  0 19&$-$68 52 19&   0.067&   0.004  \\
  5  0 19&$-$68 55 48&   0.087&   0.004  \\
  5  0 19&$-$68 45 17&   0.054&   0.003  \\
  5  0 19&$-$68 48 46&   0.061&   0.004  \\
  5  0 20&$-$68 41 45&   0.068&   0.004  \\
  5  0 20&$-$68 38 16&   0.059&   0.003  \\
\hline
\end{tabular}
\end{table}

In figure 4, locations of 1123 sections, for which  E(V$-$I) could be
reliably estimated are shown. It can be seen that only 16\% of the total number of region 
is left out. In the figure, different colours are used
to indicate different range in the reddening value. Central regions 
show low and similar reddening values. Towards the western end, some regions
were found to show higher reddening. Eastern end is found to show higher values of 
reddening, and the highest 
estimated values of reddening were found for regions located here. The most striking feature of 
this figure is that, higher reddening is found at the ends of the bar and not in the 
central regions.
\begin{figure*}
\centering
\includegraphics[width=17cm]{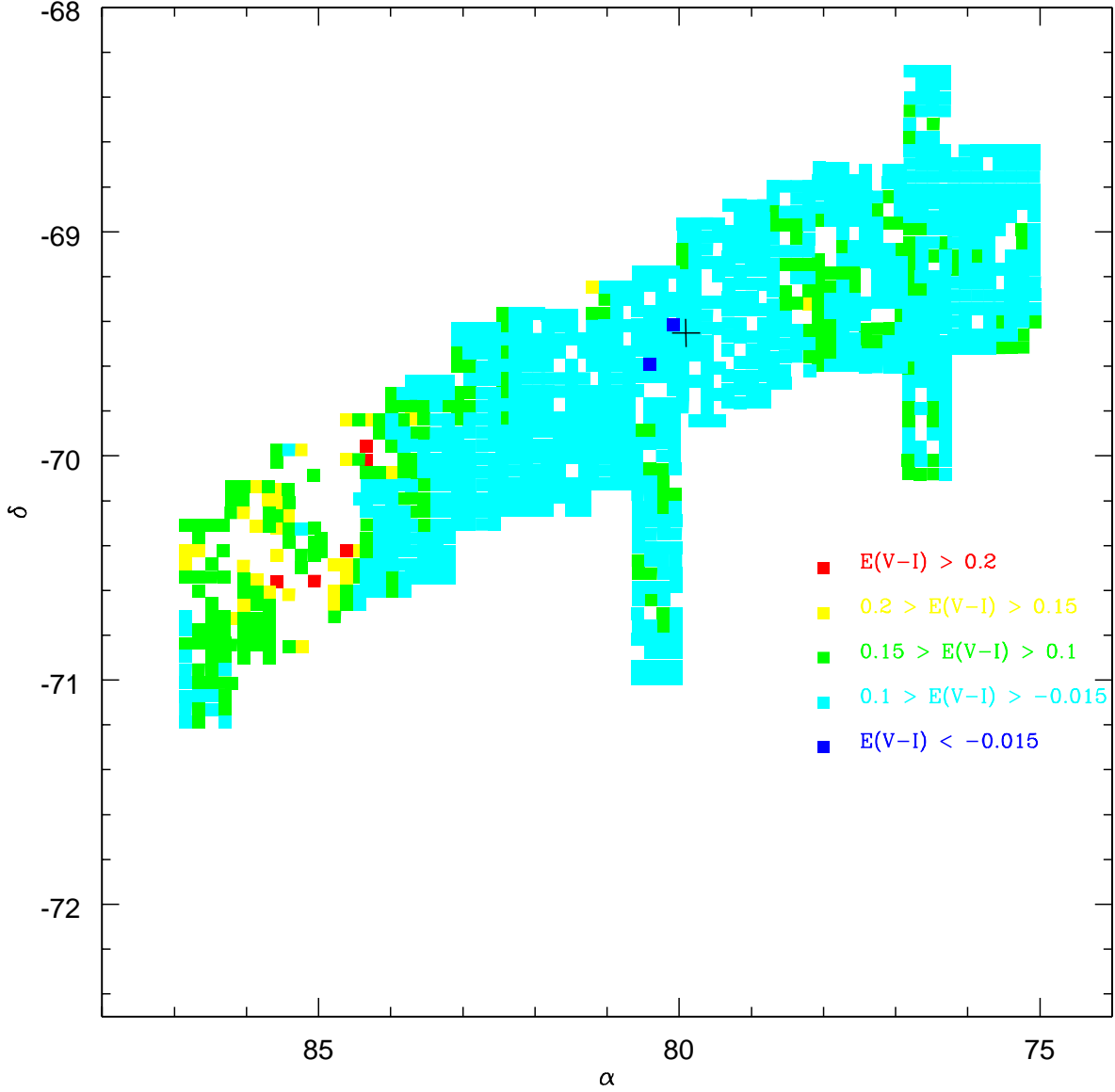}
\caption{Location of the sections for which the E(V$-$I) reddening values were estimated.
Different colours are used to denote locations with different values of reddening, which is
explained in the figure. The missing points are due to locations for which reliable
estimate of reddening could not be obtained.
 The plus sign denotes the optical center of the LMC.}
\label{figure4}
\end{figure*}

The total average reddening was also estimated and it was found to be E(V$-$I) = 
0.08 $\pm$ 0.04 mag.
This indicates that in general, reddening towards the bar regions are small. Also, the
relatively low value of dispersion indicates that the amount of differential reddening across the
face of the LMC bar is insignificant.

\begin{figure}
\resizebox{\hsize}{!}{\includegraphics{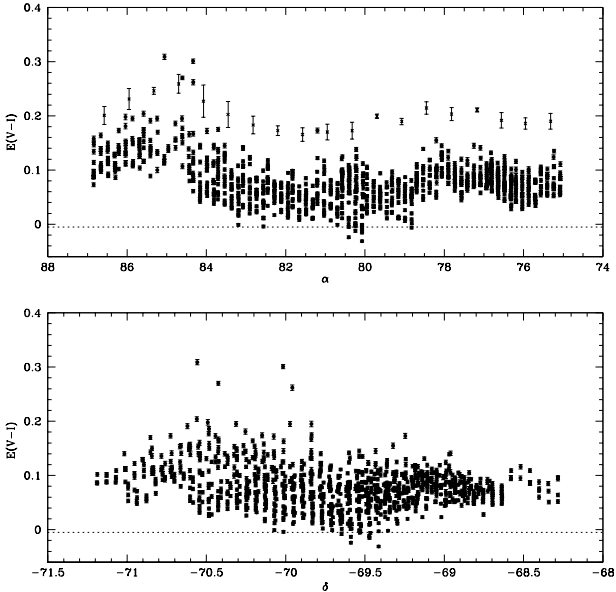}}
\caption{ E(V$-$I) reddening values are shown as a function of $\alpha$ in the
top panel and as a function of $\delta$ in the lower panel. The error in the estimation is also shown.
The crosses denote the reddening estimation on Udalski et al. (\cite{u99}), which is the average
for each strip. The error denotes the scatter in the reddening estimated for four subsections in a strip.}
\label{figure5}
\end{figure}
Reddening values are plotted as a function of $\alpha$ and $\delta$
in figure 5. It can be readily seen that, on an average, the E(V$-$I) value is 
$\le$ 0.1 mag for most of the regions. Eastern regions have slightly higher 
values of reddening upto 0.2 mag. Only a few
regions were found to have E(V$-$I) value higher than 0.2 mag. 
It can be seen that, seven locations
are found to show negetive reddening values. Of these, two locations
showed reddening values lower than $-$0.015 mag. In the statistical sense, this
is a very small fraction of the total sample. The crosses  in the figure 
denote the average reddening
values as estimated by Udalski et al. (\cite{u99}) and coverted to E(V$-$I).
The error denotes the scatter in the reddening estimates of 4 subsections 
within each strip. 
These values are considerably larger than the values estimated in this study.
This discrepancy is discussed in the following section.

\section{Results and Discussion}
Udalski et al. (\cite{u99}) presented average E(B$-$V) reddening values for 84 sections,
whereas E(V$-$I) reddening values of 1123 sections are presented here, where both
studies used the same data in the same region. Area used to estimate average
reddening in this study corresponds
to $\sim$ 46 pc $\times$ 46 pc (assuming 13.4 pc for a distance modulus of 18.24 mag). On the
other hand, the area averaged by Udalski et al. (\cite{u99}) is $\sim$ 190 pc $\times$ 190 pc.
Thus reddening values presented here have 16 times better spatial resolution.
The values presented here are thus more suited for studies which require reddening estimates
on smaller length scales.

The adopted value of (V$-$I)$_0$ = 0.92 needs some attention. This value has been
adopted from Olsen \& Salyk (\cite{os02}). They selected this value so as to 
produce a median reddening equal to that measured by Schlegel et al. (\cite{sch98}). 
Though none of the fields considered by Olsen \& Salyk (\cite{os02}) fall within the bar, 
the assumption can be assumed to be valid considering the smooth distribution of reddening in the
central region of the LMC. As 
pointed out by the referee, this may be a significant source of uncertainty in the final
reddening values. 
Statistically, only 0.18\% of regions showed negative reddening, which is a
very negligible fraction.
The fact that only two locations showed negative reddening asserts that the
assumed value of the mean colour of the red clump stars, which is the  zero-point for reddening 
estimation is not incorrect. The mean value of the intrinsic colour cannot be higher than the
value assumed here, but
at the same time, it does not rule out the possibility of a lower value.
The mean colour of red clump population in the solar neighbourhood
were derived by Paczynski \& Stanek (\cite{ps98}), and was found to be (V$-$I)$_{mean}$ = 1.01 mag.
Whereas the mean colour for the bulge population of red clump stars were found 
to be 1.22 mag. This difference was interpreted as due to the bulge population having a 
broader range and a higher value of average metallicity than the local disk population.
The fact that mean colour of red clump stars in the LMC is found to 
be bluer (0.92 mag) could be explained as due to
red clump population in the LMC being metal poor than the local disk population.
Girardi \& Salaris (\cite{gs01}) modeled the red clump population and estimated their mean (V$-$I)
and M$_I$ values as a function of age and metallicity. From their figure 1, it can be
seen that for Z=0.004, the mean (V$-$I) value is between 0.95 and 0.90 mag, for an age range
of about 2 -- 12 Gyr. This gives a theoretical support to the mean 
value of 0.92 mag assumed by Olsen \& Salyk (\cite{os02}). 
The value of mean colour was also found to
increase with metallicity. If the mean colour of red clump stars were
found to be different from the value used here, then the reddening of the locations presented 
here would shift by the value equal to the difference between the present zero-point and the 
new value. The difference in reddening between two locations would still remain the same
as estimated here.

\begin{figure}
\resizebox{\hsize}{!}{\includegraphics{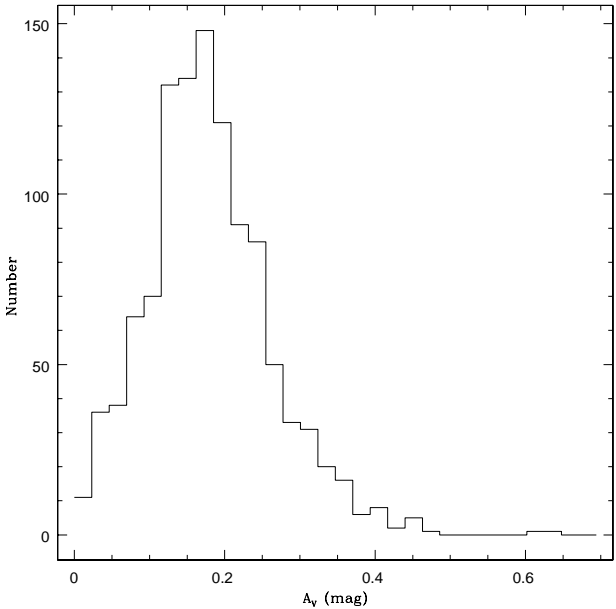}}
\caption{ A histogram of extinction, A$_V$ mag. A single peaked distribution can be noticed.
}
\label{figure6}
\end{figure}
The average value of E(B$-$V) reddening as estimated by Udalski et al. (\cite{u00}) was
0.143 mag, which would translate to E(V$-$I) = 1.4~E(B$-$V) = 0.20 mag. This is very much
higher than the value presented here. The average reddening presented in this study could be biased
towards lower value due to the fact that regions with higher reddening are not considered.
The average reddening values for each strip as estimated by Udalski et al. (\cite{u99}) is
shown in figure 5, where the E(B$-$V) values are converted to E(V$-$I) values 
using the relation, E(V$-$I) = 1.4~E(B$-$V). 
It can be seen that the values estimated here are lower than those estimated
by Udalski et al. (\cite{u99}). There seems to be more or less a constant difference between the two values 
irrespective of location.  There could be two reasons for this difference. The first reason is,
may be the zero-point used should have been lower than 0.92 mag by about 0.1 mag.
Lowering of the zero-point from 0.92 to $\sim$ 0.82 would mean that the red clump stars have a
metallicity around Z=0.001 (Girardi \& Salaris (\cite{gs01}).
Harris et al. (\cite{h97}) used around 2000 OB main-sequence stars to construct a map of the
reddening in the region observed by Zaritsky et al. (\cite{z97}). These regions are located to the
north of the region used in this study. They find a mean reddening of E(B$-$V)= 0.20 mag.
This would correspond to E(V$-$I) = 0.28 mag 
(assuming the relation, E(V$-$I) = 1.4~E(B$-$V)). 
Stanek et al. (\cite{szh98}) used red clump stars
in two specific locations within the above region. The estimated values of reddening for
these regions were E(V$-$I) = 0.20 and 0.21 mag. It has been demonstrated by Zaritsky (\cite{z99})
that the extinction property of the LMC vary both spatially and as a function of stellar
population. Recently Zaritsky et al. (\cite{z04}) presented an extinction map for two stellar
population in the LMC. These were obtained by fitting stellar atmospheric
models to a set of cool and hot stars. They found that dust is highly localised near
the younger, hotter stars, and in particular towards regions immediately east and
northeast of the center of the LMC. Also, aside from the regions of higher extinction, no
discernible global pattern were found. The reddening values estimated in the present study
 are slightly less than those presented in the above reference. 
The stellar population studied here is predominantly located in the bar, which may be
located in front of the dust and the gas clouds in the LMC. On the other hand, the
stellar population analysed by Zaritsky et al. (\cite{z04}) could belong to a younger population
and hence located in the disk, which may be mixed with gas and dust. Zaritsky et al. (\cite{z04})
found that extinction towards older stars is bimodal such that it could be modeled as
stars in front and behind a thin dust layer. For comparison, the histogram of the extinction
in V, A$_V$, is plotted in figure 6, where A$_V$ is estimated as A$_V$ = 3.24$\times$(E(V$-$I)/1.4).
It can be seen that the profile is similar to the profile presented by Zaritsky et al. (\cite{z04}),
but without the second peak. This again supports the fact that regions with higher reddening have
been selectively omitted. This can therefore skew the value of average reddening towards lower values.
The regions presented here thus represent locations which lie in front of the dust 
layer as mentioned by Zaritsky et al. (\cite{z04}). They also mention that the Galactic foreground 
extinction is estimated to be minimal, of the order of A$_V$ $\sim$ 0.05 mag. The peak of the
profile in figure 6, would shift marginally towards lower value, when the Galactic foreground
extinction is removed. Using the above relation, the E(V$-$I) reddening due to the Galactic
foreground can be estimated as $\sim$ 0.01 mag. Hence the major part of the reddening is due 
to the LMC itself.

A couple of locations near the center were found to show statistically significant
negative reddening. This may indicate that
the red clump population close to the center may be different from the rest of the 
bar region in terms of metallicity and/or age. If we consider near-zero reddening near these
locations, then the mean (V$-$I) required is 0.905 mag.  From Girardi \& Salaris (\cite{gs01}, figure 1),
the mean colour for a matallicity of Z = 0.001 was found to be less than 0.85 mag. Therefore,
for a population with similar age, the required increase in the 
metallicity would only be marginally more. If the metallicity were  to be
constant, then a red clump population with age younger than 1 --2 Gyr is required to explain the
observed colour. The above two ranges are quite possible, and hence the observed red clump  
population near the center could be either younger or slightly metal rich than the rest of the
population in the bar.

The reddening map as shown in figure 4, indicates that central part of the bar
region has very small reddening. This would mean that there is not much gas/dust present here. 
The reddening is more at the ends of the bar, with the
eastern end having more reddening than the western end. Towards the west of the bar, the 
reddening increases initially, and then decreases to the values found near the center. 
Hence  presence of gas/dust are found only for a small stretch of the western bar. 
On the other hand, many locations in the eastern end of the bar show relatively high reddening.
 The reddening between $\alpha$ = 84$^o$ -- 85$^o$ shows that a few regions have highest 
values of reddening. Further eastward of 85$^o$, the reddening is found to be 
slightly higher, but with similar dispersion
as seen in the rest of the bar. This indicates that only the net reddening is higher, 
whereas the dispersion, which corresponds to differential reddening, is not higher. 
This can be clearly seen from the middle panel of figure 3, where the value of $\sigma$ 
is not found to be higher for the eastern end, though a few points show higher values.
The top panel of figure 5, also indicates that a higher reddening is found towards
the east, similar to the result of Zaritsky et al. (\cite{z04}). We try to find the possible
reason for this increased reddening.
The HI density map presented in figure 4 of Kim et al. (\cite{k98}) shows a very large density of 
H I at $\alpha$ $\sim$ 85$^o$ and $\delta$ between 69$^o$ -- 71$^o$. This dense gas should 
have caused very high reddening in these regions. The reddening values obtained here should
then be very high, rather than the estimated small increase. 
Thus the higher reddening observed for the east, when compared to the west, can be 
explained as due to the presence of increased amount of gas and dust in the eastern region. 
On the other hand, the reddening 
observed is not too high to account for very large amount of gas as observed 
by Kim et al. (\cite{k98}). Thus we find that
only a small fraction of the H I gas seems to be located in front of the bar, giving rise to 
increased reddening. Most importantly, the
large column density of H I clouds seen in this direction is most likely to be 
located behind the bar region.

The main results of this study are summarised as follows:
\begin{description}
\item The E(V$-$I) reddening values of 1123 locations in the bar region of the LMC are 
presented here. These are estimated using the red clump stars identified from the OGLE II catalogue.
These estimations can be considered to be homogeneous with minimum systematic error.
\item The reddening values were found to have values between 0.1 and 0.3 mag, with an average value of
0.08 $\pm$ 0.04 mag, indicating a more or less uniform reddening throughout the bar region.
The average presented here is likely to be skewed towards lower values of reddening, due to selective
omission of highly reddened regions.
\item The eastern end of the bar is found to have slightly increased reddening, 
with respect to the rest of the bar. This indicates that there is some amount of gas/dust located
in front of the bar. 
On the other hand, this increased reddening is not enough to account for the
large amount of HI column density observed. Therefore,
most of the H I clouds as seen in the H I maps is likely to 
be located behind the bar.
\end{description}
 
\begin{acknowledgements}
I thank Prof. N.K.~Rao for suggesting to publish the results presented here. I also
thank the referee for bringing the best out this paper. 
\end{acknowledgements}

\end{document}